\documentclass[twocolumn,amsmath,amssymb,10pt,superscriptaddress,a4paper,letterpaper,fleqn]{revtex4-1}
\usepackage{amssymb}
\usepackage{epsfig}
\usepackage{graphicx}
\usepackage{dcolumn}
\usepackage{array}
\usepackage{bm}
\usepackage{fancyheadings}
\usepackage{longtable}
\usepackage{multirow}
\usepackage{float}
\usepackage{afterpage}
\usepackage{color}

\setlongtables
\usepackage[breaklinks=true,linkbordercolor={1 1 1}]{hyperref}

\begin{document}
\setcounter{page}{1}

%%% **********************************************************************

\title{
%% Please do not remove the line below
\qquad \\ \qquad \\ \qquad \\  \qquad \\  \qquad \\ \qquad \\
%% Change title, authors, afiliation and type  your abstract
Targets for Precision Measurements}

\author{W. Loveland}
\email[Corresponding author: ]{lovelanw@onid.orst.edu}
\affiliation{Chemistry Dept., Oregon State University, Corvallis, OR , 97331, USA}

\author{L. Yao}
\affiliation{Chemistry Dept., Oregon State University, Corvallis, OR , 97331, USA}

\author{David M. Asner}
\affiliation{Pacific Northwest National Laboratory, Richland, WA 99351, USA}

\author{R. G. Baker}
\affiliation {California Polytechnic State University, San Luis Obispo, CA, 93407, USA}

\author{J. Bundgaard}
\affiliation{Colorado School of Mines, Golden, CO, 80401, USA}

\author{E. Burgett}
\affiliation{Idaho State University, Pocatello, ID 83209, USA}

\author{M. Cunningham}
\affiliation{Lawrence Livermore National Laboratory, Livermore, CA 94550, USA}

\author{J. Deaven}
\affiliation{Idaho State University, Pocatello, ID 83209, USA}

\author{D.L. Duke}
\affiliation{Colorado School of Mines, Golden, CO, 80401, USA}

\author{U. Greife}
\affiliation{Colorado School of Mines, Golden, CO, 80401, USA}

\author{S. Grimes}
\affiliation{Ohio University, Athens, OH, 45701, USA}

\author{M. Heffner}
\affiliation{Lawrence Livermore National Laboratory, Livermore, CA 94550, USA}

\author{T. Hill}
\affiliation{Idaho National Laboratory, Idaho Falls, ID 83415, USA}

\author{D. Isenhower}
\affiliation{Abilene Christian University, Abilene, TX, 79699, USA}

\author{J. L. Klay}
\affiliation {California Polytechnic State University, San Luis Obispo, CA, 93407, USA}

\author{V. Kleinrath}
\affiliation{Idaho State University, Pocatello, ID 83209, USA}

\author{N. Kornilov}
\affiliation{Ohio University, Athens, OH, 45701, USA}

\author{A.B. Laptev}
\affiliation{Los Alamos National Laboratory, Los Alamos, NM 87545, USA}

\author{T.N. Massey}
\affiliation{Ohio University, Athens, OH, 45701, USA}

\author{R. Meharchand}
\affiliation{Los Alamos National Laboratory, Los Alamos, NM 87545, USA}

\author{H. Qu}
\affiliation{Abilene Christian University, Abilene, TX, 79699, USA}

\author{J. Ruz}
\affiliation{Lawrence Livermore National Laboratory, Livermore, CA 94550, USA}

\author{S. Sangiorgio}
\affiliation{Lawrence Livermore National Laboratory, Livermore, CA 94550, USA}

\author{B. Selhan}
\affiliation{Lawrence Livermore National Laboratory, Livermore, CA 94550, USA}

\author{L. Snyder}
\affiliation{Lawrence Livermore National Laboratory, Livermore, CA 94550, USA}

\author{S. Stave}
\affiliation{Pacific Northwest National Laboratory, Richland, WA 99351, USA}

\author{G. Tatishvili}
\affiliation{Pacific Northwest National Laboratory, Richland, WA 99351, USA}

\author{R.T. Thornton}
\affiliation{Abilene Christian University, Abilene, TX, 79699, USA}

\author{F. Tovesson}
\affiliation{Los Alamos National Laboratory, Los Alamos, NM 87545, USA}

\author{D. Towell}
\affiliation{Abilene Christian University, Abilene, TX, 79699, USA}

\author{R. S. Towell}
\affiliation{Abilene Christian University, Abilene, TX, 79699, USA}

\author{S. Watson}
\affiliation{Abilene Christian University, Abilene, TX, 79699, USA}

\author{B. Wendt}
\affiliation{Idaho State University, Pocatello, ID 83209, USA}

\author{L. Wood}
\affiliation{Pacific Northwest National Laboratory, Richland, WA 99351, USA}

%\affiliation{International Atomic Energy Agency, PO Box 100, Vienna-A-1400, Austria} 

\date{\today} 
%\received{8 March 2013; revised received XX June 2013; accepted XX September 2013}

\begin{abstract}
{The general properties needed in targets (sources) for high precision, high accuracy measurements are reviewed.  The application of these principles to the problem of developing targets for the Fission TPC is described.  Longer term issues, such as the availability of actinide materials, improved knowledge of energy losses and straggling and the stability of targets during irradiation are also discussed.
}
\end{abstract}
\maketitle

%%% DO NOT EDIT the following section enclosed by *****
%%% ***************************************************
\lhead{ND 2013 Article $\dots$}
\chead{NUCLEAR DATA SHEETS}
\rhead{A. Author1 \textit{et al.}}
\lfoot{}
\rfoot{}
\renewcommand{\footrulewidth}{0.4pt}
%%% ***************************************************

%%% EDIT: the body of your text starts here, you may use as many \section, \subsection, \subsubsection
%%% \begin{figure}, \begin{tabular} and \begin{equations} as needed. Please note that each \begin{}
%%% must be closed with the corresponding \end{} and that section titles should be in capital
%%% letters. Current text should be eventually deleted.

\section{ INTRODUCTION}

An important component of many high precision, high accuracy reaction-based measurements of nuclear properties is a suitable target. What constitutes a suitable target for a given measurement depends on the reaction being studied and the quantities being measured. For example, while most targets are solids, for some reactions, gaseous targets or even liquid targets are preferred. The preparation of suitable targets for a given measurement is an art, practiced by an ever shrinking group of scientists, largely chemists who have expertise in radio chemistry. Fortunately many of the methods employed in target making are preserved in various review articles \cite{r1,r2,r3,r4,r5} and the collected proceedings of the International Nuclear Target Development Society, many of which are published as special sections of this journal.

Most nuclear targets are freestanding solids or solids supported on a backing material. The former target form is preferred but frequently is not possible. One would like the target  backing material to be as thin as possible consistent with the mechanical stability of the target and not to undergo any nuclear reactions leading to products that interfere with the quantity being measured. For heavy element targets,  carbon,  aluminum and titanium are frequently used as target backing materials for charged particle induced reactions with polymeric materials also being used for neutron induced reactions. For highly radioactive targets it is common to deposit a thin layer of gold or nickel on the target to prevent self transfer, but one must ensure that this coating does not undergo interfering nuclear reactions or cause excessive energy loss or straggling. The elemental and isotopic composition of the target material must be well-known. Typical target thicknesses for high accuracy, high precision measurements are 50 to 500 $\mu$g/cm$^{2}$. The uniformity of the target deposit should be such as to have less than 5$\%$ variation in thickness across the face of the target. Vacuum volatilization is a relatively easy way to achieve the desired uniformity but the efficiency of deposition is usually less than 10$\%$, making it unsuitable for preparation of targets of low abundance nuclides.

\section{TARGETS FOR THE FISSION TPC}

The Fission Time Projection Chamber (TPC) \cite{HBK} is a new innovative tool for measuring neutron-induced fission cross sections.  This device should allow fission cross section measurements to be performed with 1$\%$ accuracy and because of the tracking of each event and full event reconstruction capability, it will involve different systematic errors compared to previous measurements \cite{chadwick}.  Various attempts to implement the Fission TPC idea are described in the literature \cite{fid, heffner} along with the systematic errors associated with such measurements \cite{heffner, HBK}.

The preparation of targets for the Fission TPC involves several special challenges \cite{wdljb}.  Currently we make targets as solid deposits of actinides ($^{232}$Th, $^{235}$U, $^{238}$U, $^{239}$Pu,$^{ 248}$Cm, $^{252}$Cf) on C or Al or polypropylene backings.  It has been suggested that a gaseous actinide target would be preferable. \cite{HBK}  This type of target would reduce the corrections for energy loss in backing materials and allow easy detection of both fragments from a fission event.  One must be careful that reactive or unstable gaseous compounds are not used in the TPC for obvious reasons.  If a gaseous compound of plutonium is to be prepared and used as a nuclear target, it seems clear that one of the primary candidates must be (1, 1, 1, 5, 5, 5-hexafluoro-2, 4-pentanedionato) plutonium, or more casually described as plutonium hexafluoroacetylacetonate.    It is well known that $\beta$-diketonates of this type form very stable, easy prepared complexes of all the actinide elements.  These compounds are routinely used in gas chromatography of the actinides.  At room temperature, they are liquids, prepared by solvent extraction of the metals from aqueous solution.  Typically they are vaporized in the injection cell of a gas chromatograph beginning at temperatures of about 100 $^{\circ}$C and being held at 200-300 $^{\circ}$C  in the chromatographic column.  

Despite the relative ease of target preparation, there are several severe disadvantages of the use of such gaseous compounds as targets for a TPC.  As stated above, to insure the compounds remain in the gaseous state, they must be held at 200-300 $^{\circ}$C  and all surfaces in contact with the target must also be at this temperature.  We believe this means that several critical elements of the TPC must  be able to withstand the rigors of a high operating temperature.  Based upon our previous experience of working in a storage ring where similar constraints were applied to detectors and electronics, this precludes the use of epoxy, many resins, solder, non-ceramic cables, etc.  This constraint adds to the difficulty of construction  of the device.  Perhaps the most challenging aspect of the use of a gaseous actinide target relates to the radiation safety problems posed by such a target.  The target is obviously a long-lived radioactive gas that is an alpha-emitter.  The target and  detector will have to be used in a glove box or similar enclosure with adequate attention being given to the emitted gases.  Any rupture of a gas containment  structure could have significant radiological consequences.

The target backing materials we have used to date are somewhat thicker than optimal (100 $\mu$g/cm$^{2}$ C, 540 $\mu$g/cm$^{2}$ Al and 280 $\mu$g/cm$^{2}$ polypropylene metalized with 2.4 $\mu$g/cm$^{2}$ Al).  This choice was made to be sure the targets were mechanically robust to greatly reduce the likelihood of breaking a target during the ``shake-down" phase of TPC development.  Thinner target backing materials are certainly possible but will necessitate well-established and tested procedures for dealing with broken or disintegrated targets like $^{239}$Pu.  It is not a question of whether these thinner backing targets will break but when they will break.

We have been fortunate to have actinide target materials of high isotopic purity (99.91$\%$ $^{235}$U, 99.9$\%$ $^{238}$U and nominal 99.99 $\%$ $^{239}$Pu).  We measure the uniformity of the targets we prepare by autoradiography.  For targets prepared by vacuum volatilization ($^{232}$Th, $^{235}$U, $^{238}$U) the targets show a variation of $\le$ 1.5 $\%$ in thickness over the 1 cm beam diameter.  For targets prepared by molecular plating ($^{239}$Pu) there is greater variation (10 $\pm$ 5 $\%$ ) in thickness over the 1 cm beam diameter.  However one must note that the TPC self-autoradiographs each target from the emitted decay $\alpha$-particle tracks.  Thus knowing the exact location on the target where a fission track originates allows one to correct for this variation.

The surface morphology of these targets has been extensively examined using atomic force microscopy, scanning electron microscopy, x-ray diffraction and digital optical microscopy. \cite{sadi}  The principal funding of these studies is the characterization of the chemical composition and structure of the molecular plated deposits.  The composition of the deposits is complex, does not include water molecules and probably include the presence of U(VI) in deposits of U.  

One of the unique features of the TPC approach to measuring fission cross sections is the ability to track fission events back to a specific location on the target.  (The position resolution of the TPC has been estimated to be 379 $\mu$m with a track angle resolution of 37 mRad \cite{HBK}).  Fission cross sections are frequently measured as ratios of a given cross section to that of $^{235}$U(n,f), a relatively well known quantity.  Usually this involves separate measurements of the $^{235}$U(n,f) and Actinide (n,f) cross sections with a given detector.  The beauty of the TPC is that, because of its tracking capability, one can {\bf simultaneously} measure the  $^{235}$U(n,f) and Actinide (n,f) cross sections under exactly the same conditions (beam, detector configuration, etc.).  This is done by constructing the targets to have non-overlapping wedges of different target materials.  The TPC tracks in which target segment wedge the event originated, and assigns the event to a particular reaction.  In Figure 1, we show a typical wedge target that we made to demonstrate the concept.

\begin{figure}[!htb]
\includegraphics[width=0.45\columnwidth]{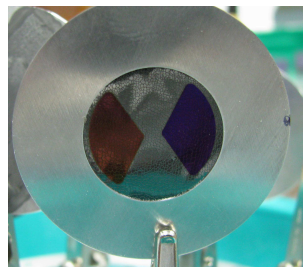}
\caption{TPC test target showing wedges of $^{232}$Th and $^{235}$U.}
\label{fig1}
\end{figure}

To be sure this scheme works, especially for the more complex targets, we have made a series of ``position resolution" targets of $^{252}$Cf where $^{252}$Cf spots are distributed in a simple  and complex patterns testing resolution at the 100 $\mu$m level in the latter case.

\section{LONGER TERM ISSUES}

As the Fission TPC moves from a concept testing phase into a true production phase, there will be more stringent demands placed upon the targets (and the target makers).  Some of these demands have been discussed above, i.e, thinner target backings, detailed information on the chemical and isotopic composition of individual targets, etc.  These improvements  are straightforward and can be met with appropriate effort.  (One should note, in fairness, that many target characterization methods are destructive and will require performing the measurements on several samples and summarizing the results by statistical methods).

There are some longer term issues that are not so easy to address.  The first of these is the availability of suitable high purity actinide materials.  We have been fortunate to get high purity actinide materials from ``private stocks" of U.S. national laboratories.  We are running out of this material and the commercially available (ORNL) material is much less pure (typically $\le$ 93 $\%$ isotopic purity) and for some materials, such as $^{248}$Cm, so expensive and in such short supply, as to be prohibitive.

``Two-sided" operation of the Fission TPC (i.e., detecting both fragments from a fission event) brings significant challenges in characterizing the fragments passing through the target backing.  Also in ``one-sided" operation, $\alpha$-particles and fission fragments leaving the target at small angles will undergo energy loss, straggling and small angle scattering.  From interactions with the actinide deposit only, it has been estimated \cite{HBK} that this will only involve $\sim$ 2$\%$ of the emitted particles and is negligible for cross section measurements.  However interactions with the target backing may be more significant, especially if the Fission TPC is used for other fission measurements besides cross sections.

Several studies \cite{jyv} have shown that the conventional solution to studying these problems, the use of the computer program SRIM \cite{SRIM} is not adequate for high accuracy, high precision studies of fission fragments.  A solution has been proposed and it involves experimentally characterizing each target/backing combination as to energy loss, straggling and scattering by use of a time of flight/energy measurement of $^{252}$Cf fission fragments passing through the target/backing combination.

The practical consequences of having to perform these detailed characterizations of the Fission TPC targets will require many months of effort by the target-maker(s).

\section{ CONCLUSIONS}

Target preparation for the Fission TPC is demanding because of the use of multi-isotopic targets, the quest for ever thinner target backings and the need for exacting characterizations of the individual targets.

%%% A blank line creates a paragraph; multiple blank lines are equivalent to one; line breaks have no effect

The authors are  indebted to Dr. John D. Baker (deceased) who participated in developing many of the ideas used preparing targets for the Fission TPC, and Dr. Chris McGrath who purified the $^{235}$UF$_{4}$ for use in this project. 
This work was funded by the Office of Nuclear Physics, Office of Science of the U.S. Department of Energy, under Contract No. DE-FG06-97ER41026.

%%% IMPORTANT: When preparing bibliography observe strictly the layout below (initials in front of the names,
%%% no names in capitals, not more than four authors, no titles, volume number in bold, year at the end within parenthesis,
%%% dot (.) at the end.

\end{document}